\documentstyle[aps]{revtex}
\begin{document}
\title{Realization of an effective
ultrahigh magnetic field on a nanoscale}
\author{S. T. Chui}
\address{ Bartol Research Institute, University of Delaware, Newark, DE
19716}
\author{Jian-Tao Wang, Lei Zhou, K. Esfarjani and Y. Kawazoe}
\address{Institute for Materials Research, Tohoku University, 
Sendai 980-8577, Japan}
\maketitle
\begin{abstract}
In tunnel junctions of which at least 
one side is a ferromagnet, very large magnetic
polarization change ($\approx 0.1 
\mu_B$) and splitting of the spin up and spin down Fermi energy ($\approx
0.1 eV$) can be created under steady state finite current conditions 
(bias voltage $\approx$
1 volt). This is {\bf much higher} than can be created by
the highest magnetic field on earth. We illustrate this with a specific
calculation of a recently observed very
large Hall effect in the Al side of a Co-I-Al tunnel 
junction. Other recent experiments that support this idea are discussed.
\end{abstract}
\section{Introduction and summary}
There is recently much interest in spin polarized transport in
ferromagnetic tunnel junctions, motivated by the high D.C. magnetoresistance
observed,\cite{Moodera,Miyazaki} and their potential applications as magnetic
sensors and nonvolatile memories. These structures are junctions of two metals
with a thin ($\approx 10 \AA\ $) 
layer of an insulator in between. They are symbolically denoted
as an F-I-F structure where I stands for insulator. 
The spin polarized transport in this system has been studied theoretically
without taking into account the electron-electron (e-e) 
interation\cite{Johnson,Son,Valet,Deweert}
and including the e-e interaction.\cite{elecpp,others,bias}
It is found that\cite{elecpp} in tunnel junctions
at least one side of which is a ferromagnet, under
steady state moderate currents ($\approx 10^{-5} $ Ampere) 
there will be a {\bf very large}
change in the polarization ($\approx 0.1 \mu_B$)
and splitting of the spin up and spin down Fermi energies ($\approx 0.1 eV$).

The implication of this splitting has not been explored.
By applying different voltages, the Fermi levels can be moved and
different parts of the band
structure can be probed. Different physical characteristics of different
parts of the band structure can be manifested.
One can envision inducing different phase transitions
by tuning the applied voltage.

Experimental results that illustrate the new physics that can happen
as a result of the splitting are beginning to appear.
The big splitting of the spin up and spin down Fermi energies 
provides for an explanation of the bias dependence that is
experimentally observed.\cite{bias}
We discussed here an illustration of the probing of a different part 
of the band structure with a detailed calculation of a giant anomolous
Hall resistivity recently discovered in Co-I-Al
tunnel junctions.\cite{GHE} Good agreements with
the experimental results are obtained. We then conclude by discussing two 
other experiments that provide further illustration
for the conclusion discussed here.

The changes in the Fermi levels can in principle also be produced by external
magnetic fields of strength of the order
of 1000 Teslas. This field strength is much higher than can
be created by the highest magnetic field on earth.  This presents a 
possibility of a new area of research of ultrahigh magnetic field physics.

That a current can induce a magnetization in a ferromagnetic-
paramagnetic junction (no insulator in between) was first discussed by
Johnson and Silsbee\cite{Johnson} and van Son et al.\cite{Son}. 
The magnetized current from the ferromagnet induces a magnetization
in the paramagnet due to a spin bottleneck effect
and causes a splitting of the spin up and the spin 
down Fermi energies $\Delta \mu$. The ratio of this splitting to the driving 
current $I$ is of the order of the resistance of the {\bf metal}.
This same reasoning can also be applied to the F-I-P structure with an 
insulator in between, and the magnetization and the splitting
is of the same order of magnitude.
The resulting magnetization from this ``spin accumulation effect'' is
very small and is different from the physics discussed here.
For the F-I-P structure and our mechanism, the ratio $\Delta \mu /I$ is 
of the order of the resistance of the {\bf insulating barrier}.
This is several orders of magnitude larger than that expected from the
spin accumulation effect.

\section{Underlying framework}
We first recapitulate briefly the reasoning that led to the splitting of the
Fermi levels.
The physics can be approximatly described by the
following sets of equations.
The first is that of global charge current conservation:\cite{eq1}
\begin{equation}
\nabla \cdot J=-\partial \rho/\partial t
\end{equation}
Here $J$, $\rho$ are the total (including both spins)
current and charge densities. The second equation
is the diffusion equation which expresses the current density for spin s,
$J_s,$ as a sum of (a) the external driving current $J_{0s}=\sigma_sE$
that is controlled by the conductivity $\sigma_s$ and the external
electric field $E$;
(b) the current driven by a density gradient (diffusion) expressed as a
gradient of the chemical potential $\nabla \mu_s=\nabla \rho_s/N_s$ 
through the density of states at the Fermi surface $N_s$
and (c) the current driven by the internal electric ``screening'' field
$\nabla W_0$:\cite{eq2}
\begin{equation}
J_s=\sigma_s[\nabla \mu_s -\nabla W_0+eE]/e
\end{equation}
Here $\sigma_s$ is the conductivity for spin channel s.
$W_0(r)$ is the local electric (screening) potential due to the other
electric charges that is determined self-consistently:
$W_0(r)=\int d^3r'U(r-r')\rho(r')$ where $U$ is the Coulomb potential.
Because of this self-consistent screening, charge fluctuation dies off with 
a length scale $\lambda$ 
called the screening length,\cite{detailc} which is of the order of 10\AA\ 
for metals. Finally, the magnetization density {\bf changes} $M$ 
relaxes with a length scale of ${\bar l}_{sf}$
where the bar indicates a renormalization effect due to the electric 
field:\cite{elecpp}
\begin{equation}
\nabla^2 M- M/{\bar l}_{sf}^2=0
\label{x1}
\end{equation}
For the systems of interest, the spin diffusion length
${\bar l}_{sf}$, is of the order of 1000 to 10000 \AA\ and is much
larger than the screening length $\lambda$. Because of these two very 
different length scales, there is a delicate balance in the current
from the charge and the magnetization diffusion that has to be maintained.

One can solve these sets of equations on each side of the junction.
The currents on opposite sides of the junction are related by the
boundary condition that the current $J_s$ is continuous across the junction
and that
\begin{equation}
\Delta \mu_s-\Delta W=r(1-s\gamma)J_s
\end{equation}
where $r(1-s\gamma)$ is the resistivity of the junction for spin 
$s=\pm 1$.

%
%
The final solution can be developed in increasing powers of
$\lambda/{\bar l}_{sf}$.
Solving equations (1-3) for a junction of thickness d at the origin, we 
find ``dipole layers''  of charge ($\rho$) and magnetization ($M$)
densities that peaks at the junction
and dies off exponentially away from the junction
with a functional form
$f(z,l)=\exp[-(|z|-d/2)/l]$:\cite{detaila}
$$\rho\approx \rho_{1}f(z,\lambda)\lambda/{\bar l}_{sf}+\rho_2f(z,{\bar l}_{sf})
\lambda^2/{\bar l}_{sf}^2$$
$$M\approx M_0f(z,{\bar l}_{sf}).$$
In this paper, we use units of $M$ in terms of the
Bohr magneton.
The charge and magnetization densities are now coupled:
$\rho_{10}=G\rho_{20}$ where $G$ is a coefficient of the order of 
unity.\cite{detail3}
$\rho_{20}=M_0D_M/D_D.$ Here $D_M$ ($D_D$) is the diffusion constant for the 
magnetization (charge) that can be expressed in terms of the
conductivities.\cite{detaild}
The charge dipole layer densities are smaller than the magnetization
densities by a factor $\lambda/{\bar l}_{sf}$! 
This comes from the two very different length scale of change for the charge
and the magnetization degrees of freedom. There is a small correction
to the charge density that is of much
longer range than the screening length.
In general the constants $\rho_{i}$, $M_0$ on opposite sides of the 
junction can be different from each other. The {\bf total} charge on opposite
sides of the junction are opposite in sign and equal in magnitude, of course.

After matching the boundary condition (eq. (4)) one obtains
for the magnetization and the Fermi energy splitting:\cite{detaile}
\begin{equation}
M_0=FrJ
\end{equation}
\begin{equation}
\mu_+-\mu_-=0.5 M(1/N_++1/N_-)
\end{equation}
where $F$ is a coefficient of the order of the inverse density of
states at the Fermi energy\cite{detail1}.
It is different for parallel and antiparallel alignment of the
magnetizations on opposite sides of the junction.
Thus the magnetization density induced is controlled by the resistance
of the {\bf junction}, not of the metal!
To illustarate the application of this, we first
calculate the Hall conductivity
$\sigma_{xy}$ of Al in a Co-I-Al junction.

\section{Anomlous giant Hall resistivity}
Otani et al.\cite{GHE} recently measured the Hall coefficient of Al
in a Co-I-Al structure. They found that above a threshold voltage
of about 1 volt, there is a rapid increase in the Hall voltage that is of the
same sign (negative) as but much larger than the ordinary Hall 
coefficient of Al. This effect disappears as the temperature is increased 
from 30K to 150 K.
Their original explanation in terms of an extraordinary Hall effect is 
inapplicable with a revised estimate of the Hall resistivity that is several
orders of magnitude higher.

In our picture, as the external voltage $V$ is increased, the spin up and spin 
down Fermi energies will move apart by an amount $2\Delta$ proportional to $V$. 
The sample is polycrystalline experimentally\cite{GHE},
Without detail knowledge of the nature of this polycrystalline sample,
for simplicity
we  consider the crystal structure as a single fcc 
crystal of bulk Al with lattice parameter 4.0491\AA,
and  have taken the direction of
the magnetization parallel to the $z$ symmetry axis of Al.
We have calculated the Hall conductivity $\sigma_{xy}$ from first principles
as a function of $\Delta$.

Our calculations are performed using
the band structure obtained with 
the self-consistent full-potential linearized augmented-plane-wave
(FLAPW) method \cite{WIEN} under the generalized
gradient approximation (GGA) \cite{gga} with spin-orbit
coupling\cite{detailf,Singh,Novak}.
The conductivity is calculated using the Kubo formula.
The Brillouin zone sampling is performed using {\it 4000} special k-point
meshes, which yielded {\it 315} points in the irreducible Brillouin
zone.\cite{detailb}
The dependence of conductivity on the frequency have been tested on the
Kerr effect of bcc Fe.
For an inverse lifetime $\Gamma =0.45\sim 0.65eV$, our result is
in very good agreement with the experimental results. The Hall conductivity
is the zero frequenct limit of the conductivity.

Our result for the Hall conductivity
is shown in Fig. 1. The sign 
and the shape (including the threshold) are in good agreement with experiment.
The physics of the threshold and the temperature dependence can be 
understood as follows.
The band structure of Al with the spin up and spin down 
Fermi levels is shown in Fig 2 for the
case $\Delta=0.2 eV$. Two band crosses along the direction XZW. The crossing 
is spin-orbit split. At zero magnetization, this corssing is below the Fermi 
energy. When the spin-up and the spin down Fermi levels move apart, one of 
them will move towards this crossing and a significant interband contribution
will come in for the Hall conductivity. The thresold is determined by
the distance between the 
energy at this crossing and the Fermi energy at zero field.
Because the spin-orbit band gap is very small, we expect the interband
contribution will decrease significantly when the temperature becomes
comparable to this band gap.

The magnitude of our result depends on the scattering time $\tau=\hbar/\Gamma$.
Our result is in the range of 1 to 10 $1/\Omega - cm$. Because of the 
nonuniformity of the current flow due to the geometry of
the experimental sample, and becuase we do not know the detail material 
characteristic of the polycrystalline experimental sample, we do not expect
to get {\bf exact} quantitative agreement with the experimental results.
If we pick a typical $\Gamma=0.25 eV$, our result is in good agreement 
with experimentally revised value of
$\rho_{xy}=1.5 \times 10^{-8}\Omega -m$ 
for an average $\rho_{xx}=3.4\times 10^{-6}\Omega \ m$.
(Recall that $\sigma_{xy}=\rho_{xy}/\rho_{xx}^2$.)
This diagonal resistivity is consistent 
with recent unpublished results by the same authors
for a ohmic (F-P) Co-Al junction.

Our picture can be further tested experimentally. In our calculation, the Hall
resistance goes through a maximum and 
decreases as the bias voltage is further increased.
So far the experiment was carried out up to a bias when the maximum
of $\rho_{xy}$ was reached.
We predict that the Hall resistivity will decrease as the voltage is further
increased.
We next discuss some other recent experimental results that supports
the present picture.

Moodera and coworkers recently found a fascinating series of bias 
dependences, including a negative junction magnetoresistance,
for a F-I-P-F structure with varying thickness for the
paramagnet.\cite{Mod1} To explain their result, they assuumed a model
with different spin up and spin down Fermi energies. The result discussed
here provides for a possible justification of the model that they have used.

Ono, Shimada and Ootuka\cite{Ono} recently studied the magnetoresistance
of ferromagnetic single electron transistors made from tunnel structures. 
Depending on the gate voltage,
this device can be in an on or an off state. In the on-state, they found 
a magnetoresistance ratio of about 4 \%. They observed a ten fold increase
of the magnetoresistance ratio in the off-state.
In the off-state, the transport depends {\bf exponentially} on the activation
barrier which is a {\bf sum} of the Coulomb charging energy and the difference
of the Fermi energy on the left and the right hand side of the 
junction.\cite{Inoue} The large magnetoresistance can be understood in 
the present picture from the different splitting of the Fermi levels
between the cases of the parallel and the antiparallel alignment 
of the magnetizations.

In conclusion, we propose that magnetic tunnel junction provides for 
new opportunities to observe high magnetic field physics. Theoretical and 
experimental evidences are provided to support this claim.

STC thanks the hospitality of the Institute of Material Research
of Tohoku University where this work was started. We thank Professors
Y. Otani, K. Takanashi, Y. Ootuka for helpful conversation.
JTW thanks the JSPS for financial support.

\begin{figure}
\caption{
Dependence of conductivity and the spin polarization on the 
spin-up spin-down Fermi level splitting
2$\Delta$ = $E_{F}^{up}-E_{F}^{dn}$.
}
\end{figure}
\begin{figure}
\caption{
Energy bands for fcc Al in the ferromagnetic ground state with Fermi
level splitting $2\Delta= 0.4$eV. The Fermi levels are shown by 
the solid horizontal lines.}
\end{figure}

\begin{references}
\bibitem{Moodera} J. S. Moodera, L. R. Kinder, T. M. Wong and R. Meservey
Phys. Rev. Lett. 74, 3273 (1995).
\bibitem{Miyazaki} T. Miyazaki and N. Tezuka, Jour. Mag. Mag. Mat. 139
L231 (1995).
\bibitem{Johnson} M. Johnson and R. H. Silsbee, Phys. Rev. Lett. 55, 
1790 (1985); M. Johnson, Phys. Rev. Lett. 70, 2142 (1993). 
\bibitem{Son} P. C. van Son, H. Van Kempen and P. Wyder, Phys. Rev. Lett.
58, 2271 (1987)
\bibitem{Valet} T. Valet and A. Fert, Phys. Rev. B48, 7099 (1993). 
\bibitem{Deweert} 
M. J. Deweert and S. Girvin, Phys. Rev. B37, 3428 (1988). This paper
did not take into account the spin accumulation effect,
which is manifested mathematically as the continuity of the current as it
\bibitem{elecpp} S. T. Chui, Jour. Appl.  Phys. 80, 1002 (1996),
US patent no.  5757056 (1998) .
\bibitem{others}
See also
S. T. Chui Phys. Rev. B52, R3832 (1995);
S. T. Chui and J. Cullen, Phys. Rev. Lett. 74, 2118 (1995).
\bibitem{bias}
S. T. Chui, Phys. Rev. B55, 5600 (1997).
\bibitem{GHE} Y. Otani, T. Ishiyama, S. G. Kim and K. Fukamichi,
Jour. Appl. Phys. 87, 6995 (2000).
\bibitem{eq1} This corresponds to eq. (4) of ref. (3).
\bibitem{eq2} This corresponds to eq. (1) of ref. (3).
\bibitem{detaila}
From Gauss's law, the self consistent internal electric field is
$\nabla W_0\approx 4\pi \lambda (\rho_1f(z,\lambda)+\rho_2f(z,{\bar l}_{sf})
\lambda/{\bar l}_{sf}).$
\bibitem{detail3}
$G=-(\sigma^2/(\chi_0^2D_M)-b\sigma/\chi_0)
\rho_{20}/[(D_M-b\sigma)].$ 
$b=(\sigma_+-\sigma_-)/(\sigma_++\sigma_-).$
$\sigma=0.5 \sum_s\sigma_s$
\bibitem{detailc}
The screening length is $\lambda=\sqrt{1/4\pi \chi_0 e^2}$
where the effective density of state
$\chi_0=0.5(\sum_s\sigma_s)/[0.5\sum_s \sigma_s/N_s]$.
\bibitem{detaild}
$D_M=0.5\sum_s s\sigma_s/(eN_s).$
$D_D=0.5\sum_s \sigma_s/e\chi_0$,
\bibitem{detaile} For the systems that we look at, the $\Delta W$
term is much smaller and neglected.
\bibitem{detail1}
$F=[(b-\gamma)/{\bar N}_{D}-(1-\gamma b)/{\bar N}_{S}]
4/[1/N_{SA}N_{DS}-1/N_{DA}N_{SS}];
$
$1/N_{S(D)}=1/N_+\pm 1/N_-$, $f_{S(A)}=f^>\pm f^<$ for any function f.
The superscript
$<$, $>$ indicates the left and the right side of the insulator.
is a measure of the asymmetry of the conductivity between the
spin up and the spin down band.
Note that because the spin up and spin down Fermi energies are different,
the densities of states at the Fermi energies of the spin up and spin down
electrons in Al will be different.


\bibitem{WIEN}
P. Blaha, K. Schwarz,  and J. Luitz, WIEN97
(Vienna University of Technology, 1997).
Improved and updated Unix version of original copyrighted WIEN-code
by P. Blaha, K. Schwarz, P. Sorantin, and S. B. Trickey,
in Comput. Phys. Commun. {\bf 59},  399 (1990).
                                                   
\bibitem{gga}
J. P. Perdew, J. A. Chevary, S. H. Vosko, K. A. Jackson, M. R. Pederson,
D. J. Singh, and C. Fiolhais,  Phys. Rev.  B {\bf 46}, 6671 (1992);
J. P. Perdew, K. Burke, and M. Ernzerhof,
 Phys. Rev. Lett.  {\bf 77}, 3865 (1996).
\bibitem{detailf}
Spin-orbit interactions can be considered via a second variational step
 using the scalar-relativistic eigenfunctions as basis.

\bibitem{Singh}
D. Singh, {\sl Plane waves, pseudopotentials and the LAPW method},
(Kluwer Academic, 1994)

\bibitem{Novak}
P.  Novak, to be published.
\bibitem{detailb}
A detailed discussion is reported by 
P. M. Oppeneer, T. Maurer, J. Sticht, and. J K\"ubler
in Phys. Rev. B {\bf45}, 10924 (1992).
Here, we first calculate the imaginary part from the momentum sum,
the real part of conductivity $\sigma_{xy}(\omega)$ is obtained
by performing a Kramers-Kronig transformation as
\begin{equation}
{\sigma_{xy}^{(1)}(\omega) }
={2\over\pi}P\int_0^{\infty}{x\over {x^2-\omega^2} } \sigma_{xy}^{(2)}(x)dx
\end{equation}
where, $\sigma_{xy}^{(2)}$ is the imaginary part of the $\sigma_{xy}(\omega)$.
The integral are performed up to 2.0 Ry,
and about $10000$ $\omega$ mesh is used.
The static conductivity shown in Fig. 1 corresponds to $\omega=0$.
\bibitem{detail2}
From figure 3 of ref. 6, we get a Hall resistance of about 0.3 Ohm
for a current of $10^{-5}$ Ampere. 
To get the resistivity, we have to multiply by an area A and divide by a
length l. For A, we took $300 nm \times 0.5 \mu m$. The first factor is the
thickness that was quoted. The second we estimate from the picture.
For l, we took the length of the Hall bar which we estimate to be 3 micron.
From these, we get a Hall resistivity $\rho_{xy}\approx 1.5 \times
10^{-8} Ohm-m$. This is much less than the  $1.8 \times 10^{-17} Ohm-m$
in the paper.
\bibitem{Mod1} J. Moodera, J. Nowak, L. R. Kinder, P. M. Tedrow, R. J. M. van
de Veerdonk, B. A. Smits, M. vam Kampen, H. J. M. Swagten and W. J. M. de Jonge,
Phys. Re. Lett., 83, 3029 (1999).
\bibitem{Ono} K. Ono, H. Shimada and Y. Ootuka, Jour. Phys. Soc.
Jpn., 66, 1261 (1997).
\bibitem{Inoue} J. Inoue, private communication.
\end{references}
\end{document}